\documentstyle[preprint,aps]{revtex}

\input epsf.tex
\def\DESepsf(#1 width #2){\epsfxsize=#2 \epsfbox{#1}}
\begin{document}
\preprint{\vbox{\hbox{OITS-641}\hbox{COLO-HEP-389}}}
\draft
\title {A Critical Study of  B Decays to Light Pseudoscalars}
\author{ N. G. Deshpande $^{1}$, B. Dutta $^{1}$, and Sechul Oh $^{2}$\\
$^{1}$ Institute of Theoretical Science, University of Oregon, Eugene, OR
97403\\$^{2}$ Department of Physics, University of Colorado, Boulder, CO 80309}
\date{October, 1997}
\maketitle
\begin{abstract} Motivated by the large branching ratios observed for the
process $B\rightarrow\eta^{\prime}K$, we examine critically all the ingredients
that go into estimates of B decays into two light pseudoscalars. Within
factorization approximation, we examine several assumptions on the input
parameters that could have a strong bearing on the predictions. Among these are
(i) the QCD scale
$\mu$ (ii) value of the form factors (iii) value of the light quark masses, and
in particular $m_s$ (iv) the value $\xi=1/N_c$, (v) charm content of
$\eta^{\prime}$. It is possible to account for all the data without invoking new
physics, though future experiments will provide tighter constraints on the
parameter space. We find that CP violating asymmetries are in the observable
range for some modes.
\end{abstract}
\newpage \section{Introduction}

Recent CLEO measurement for the branching ratio of $B\rightarrow\eta^{\prime} K$
\cite{cleo0}
 is larger than expected. This result has initiated numerous investigations,
with some even suggesting new physics. In this paper we attempt to explain the
whole set of known results on two body decays of $B$ mesons into light
pseudoscalars within the context of Standard Model (SM) using generalized
factorization technique. This technique is very successful in decays of B meson
to D mesons \cite{stech0}. If this approximation is able to explain all two body
$B$ decays, we will have a powerful tool to extract various parameters like the
Cabibbo-Kobayashi-Maskawa (CKM) elements.  

Present attempts to explain the large branching ratio
$BR(B^{\pm} \rightarrow \eta^{\prime} K^{\pm})$ involve different assumptions.
\cite{a,b,c,d}.   Some  \cite{a,b} explain it on the basis of large form
factors, but SU(3) constraints on the form factors have been ignored. For
example, in the flavor SU(3) limit, there are relations among the form factors:
$F^{B
\rightarrow
\eta^{\prime}}(0) = (\sin\theta / \sqrt{6} + \cos\theta / \sqrt{3}) F^{B
\rightarrow \pi^{-}}(0)$ and 
$F^{B \rightarrow \pi^{-}}(0) = F^{B \rightarrow K^{-}}(0)$  (where $\theta$
denotes the $\eta - \eta^{\prime}$ mixing angle ).  Taking $F^{B
\rightarrow \eta^{\prime}}$ large could have the undesirable effect of increasing
$B\rightarrow \pi K$ and $B\rightarrow \pi\pi$ rates above the present bound.
Others have invoked charm for $\eta^{\prime}$, with contribution arising from
$b\rightarrow s(\bar c c)\rightarrow s\eta^{\prime}(\eta)$. Explanations have
been proposed with large $|f^{(c)}_{\eta^{\prime}}| \approx$ 50 MeV \cite{d}
and  
 relatively smaller value of  
$|f^{(c)}_{\eta^{\prime}}| \approx$ 6 MeV \cite{c}. The effect of low strange
quark mass in enhancing the rate has been noted \cite{a,c}. In an interesting
paper \cite{ros}, consequences of large $B\rightarrow \eta^{\prime}K$ branching
ratio from purely SU(3) viewpoint has been studied. We shall focus our attention
on a more dynamical analysis based on generalized factorization in the spirit of
Ali and Greub \cite{c}.

The branching ratio of $B\rightarrow\eta^{\prime} K$ depends on a number of
parameters. These parameters include the value of the strange quark mass $m_s$,
possibility of QCD scale dependence $\mu$, the size of the form factors,  the
value of the parameter 
$\xi\equiv 1/N_c$ which arises in the generalized factorization model, the
$\eta-\eta^{\prime}$ mixing angle
$\theta$, the value of the CKM elements and weak phases. We approach the problem
by first studying $B\rightarrow \pi \pi$ decays. These decays have only slight
$\mu$ dependence, and already limit the size of the form factors.  By studying
$B\rightarrow K\pi $ next, we again see the $\mu$ dependence in Wilson
coefficients (WC's) is offset by the scale dependence of $m_s$, and the branching
ratios have very slight $\mu$ dependence. It is possible to enhance
$B\rightarrow\eta^{\prime}K$ by choosing a small value $\xi$. Study of the ratio
of $B\rightarrow\eta^{\prime} K$ to  $B\rightarrow\pi K$, which is independent
of the form factors reveals that small value of $\gamma$, the weak phase, is
preferred. We are able to account for all data without assuming charm content of
$\eta^{\prime}$. With the parameter space obtained, we look at the CP
asymmetries as a function of $\gamma$, and point out that
$B\rightarrow\pi K$ and $B\rightarrow\eta K$ provide two interesting modes with
significant asymmetries.

We organize this work as follows. In section II we obtain the Wilson
coefficients  and the strange quark mass at the scale $m_b$ and $m_b/2$. In
section III we discuss the factorization approximation, in section IV first we
discuss the decays of B into
$\pi\pi$ modes. Then we discuss  $\pi K$,
$\eta^{\prime} K$ and
$\eta K$ and show the parameter space where the calculated branching ratio of 
$B\rightarrow\eta^{\prime} K$ is experimentally allowed. In section V, we
discuss the CP asymmetries in the B decay modes. Finally in section VI we
summarize our results.
 
\section{Determination of the effective Wilson coefficients}  The effective weak
Hamiltonian for hadronic $B$ decays can be written as 
\begin{eqnarray}
 H_{\Delta B =1} &=& {4 G_{F} \over \sqrt{2}} [V_{ub}V^{*}_{uq} (c_1 O^{u}_1
+c_2 O^{u}_2) 
   + V_{cb}V^{*}_{cq} (c_1 O^{c}_1 +c_2 O^{c}_2)
   - V_{tb}V^{*}_{tq} \sum_{i=3}^{12} c_{i} O_{i}] \nonumber \\ 
  &+& h.c. ,
\end{eqnarray}  where $O_{i}$'s are defined as 
\begin{eqnarray}
 O^{f}_{1} &=& \bar q_{\alpha} \gamma_{\mu} L f_{\beta} \bar f_{\beta}
\gamma^{\mu} L b_{\alpha} ,
  \ \  O^{f}_{2} = \bar q \gamma_{\mu} L f \bar f \gamma^{\mu} L b ,          
\nonumber \\ 
 O_{3(5)} &=& \bar q \gamma_{\mu} L b \Sigma \bar q^{\prime} \gamma^{\mu} L(R)
q^{\prime} ,   
  \ \ O_{4(6)} = \bar q_{\alpha} \gamma_{\mu} L b_{\beta} \Sigma \bar
q^{\prime}_{\beta} 
       \gamma^{\mu} L(R) q^{\prime}_{\alpha}  ,                        
\nonumber \\ 
 O_{7(9)} &=& {3 \over 2} \bar q \gamma_{\mu} L b \Sigma e_{q^{\prime}} \bar
q^{\prime} 
       \gamma^{\mu} R(L) q^{\prime} ,        
  \ \ O_{8(10)} ={3 \over 2} \bar q_{\alpha} \gamma_{\mu} L b_{\beta} \Sigma
e_{q^{\prime}} 
       \bar q^{\prime}_{\beta} \gamma^{\mu} R(L) q^{\prime}_{\alpha} \;,    
\nonumber \\  O_{11} &=&{g_{s}\over{32\pi^2}}m_{b}\bar q \sigma_{\mu \nu}
RT_{a}b  G_{a}^{\mu \nu} \;,\;\; O_{12} = {e\over{32\pi^2}} m_{b}\bar q
\sigma_{\mu \nu}R b  F^{\mu \nu} \;,
\end{eqnarray}  where $L(R) = (1 \mp \gamma_5)/2$, $f$ can be $u$ or $c$ quark,
$q$ can be $d$ or $s$ quark,  and $q^{\prime}$ is summed over $u$, $d$, $s$, and
$c$ quarks.  $\alpha$ and $\beta$ are  the color indices.  
$T^{a}$ is the SU(3) generator with the normalization $Tr(T^{a} T^{b}) =
\delta^{ab}/2$.  
$G^{\mu \nu}_{a}$ and $F_{\mu \nu}$ are the gluon and photon field
strength. $c_i$s  are the WC's.  
$O_1$, $O_2$ are the tree level and QCD corrected operators.  
$O_{3-6}$ are the gluon induced strong penguin operators.  $O_{7-10}$ are the
electroweak penguin  operators due to $\gamma$ and $Z$ exchange, and ``box''
diagrams at loop level. In this work we shall take into account the
chromomagnetic operator $O_{11}$ but neglect the extremely small contribution
from $O_{12}$.

We obtain the  $c_i(\mu)$s by solving the following renormalization group
equation (RGE):
\begin{eqnarray} ( -{\partial \over \partial t} + \beta (\alpha_s) {\partial
\over \partial
\alpha_s}) {\bf C}(m_W^2/\mu^2, g^2) = {\hat \gamma^T(g^2)\over 2} {\bf C}(t,
\alpha_s(\mu),\alpha_e)\,
\end{eqnarray} where $t\equiv ln(M_W^2/\mu^2)$ and  ${\bf C}$ is the column
vector consists of 
$(c_i)$s. The beta and the gamma are given by:
\begin{eqnarray}
\beta (\alpha_s) &=& -(11-{2\over 3}n_f){\alpha_s^2\over 16\pi^2} - (102-{38\over
3}n_f) {\alpha_s^4\over (16\pi^2)^2} + ...\;,\nonumber\\
\hat \gamma(\alpha_s)&=&(\gamma^{(0)}_s+
\gamma^{(1)}_{se}{\alpha_{em}\over{4\pi}}){\alpha_s \over 4\pi} +
\gamma_e^{(0)} {\alpha_{em}\over 4\pi}+
\gamma^{(1)}_s{\alpha^2_s\over(4\pi)^2} +... \;,
\end{eqnarray} where $\alpha_{em}$ is the electromagnetic coupling and
$n_f$ is the number of active quark flavours.

The anomalous-dimension matrices $\gamma^{(0)}_s$ and $\gamma^{(0)}_e$ 
determine the leading log  corrections and they are renormalization scheme
independent. The next to leading order corrections  which are determined by
$\gamma^{(1)}_{se}$ and $\gamma^{(1)}_{s}$ are renormalization scheme dependent.
The
$\gamma$'s have been determined in the reference \cite{{A},{B}}.
 
We can express
$C(\mu)$ (where $\mu$ lies between $M_W$ and $m_b$) in terms of the initial
conditions for the evolution equations :
\begin{eqnarray} C(\mu)&=&U(\mu,M_W)C(M_W).
\end{eqnarray}
$C(M_W)$s are obtained from  matching the full theory to the effective theory at
the
$M_W$ scale \cite{{B},{C}}. The WC's so far obtained are renormalization scheme
dependent. In order to make them scheme independent we need to use a suitable
matrix T \cite{B}. The  WC's at the scale $\mu=m_b$ are given by :
\begin{eqnarray} {\bar C}(\mu)&=&TU(m_b,M_W)C(M_W).
\end{eqnarray} The matrix T is given by
\begin{eqnarray} T&=&{\bf 1}+ {\hat r}^T_s {\alpha_s\over {4\pi}}+{\hat r}^T_e
{\alpha_e\over{4\pi}},
\end{eqnarray} where ${\hat r}$ depends on the number of up-type quarks and the
down type quarks, respectively. The r's are given in the references\cite{B}. In
order to determine the coefficients at the scale
$\mu<m_b$, we need to use the matching of the evolutions between the scales
larger and smaller than the threshold. In that case in the expression for T we
need to use
$\delta {\hat r}$ instead of  ${\hat r}$, where $\delta {\hat
r}=r_{u,d}-r_{u,d-1}$, where u and d are the number of up type quarks and the
number of down type quarks, respectively. The matrix elements ($O_i's$) are also
needed to be have one loop correction. The procedure is to write the one loop
matrix element in terms of the tree level matrix element and to generate the
effective Wilson coefficients
\cite{D}.
\begin{eqnarray} < c_i O_i> = \sum_{ij} c_i(\mu) [\delta_{ij} +{\alpha_s\over
4\pi}m^s_{ij} +{\alpha_{em}\over 4\pi}m^e_{ij}] <O_j>^{tree}\;.
\end{eqnarray}

\begin{eqnarray} \left(\matrix{c^{eff}_1\cr
                               c^{eff}_2\cr
                               c^{eff}_3\cr
                               c^{eff}_4\cr
                               c^{eff}_5\cr
                               c^{eff}_6\cr
                               c^{eff}_7\cr
                               c^{eff}_8\cr
                               c^{eff}_9\cr
                               c^{eff}_{10}}\right)&=
                               &\left(\matrix{\bar c_1 \cr
                                              \bar c_2 \cr
                                              \bar c_3-P_s/3 \cr
                                              \bar c_4 +P_s \cr     
                                              \bar c_5 - P_s/3 \cr
                                              \bar c_6 + P_s \cr  
                                              \bar c_7 +P_e \cr
                                              \bar c_8 \cr
                                              \bar c_9 +P_e \cr
                                              \bar c_{10} }\right),
\end{eqnarray}
 where   
 $P_s = (\alpha_s/8\pi)\bar c_2 [{{V_{cb} V_{cq}^*}\over {V_{tb} V_{tq}^*}}(10/9
+G(m_c,\mu,q^2))+{{V_{ub} V_{uq}^*}\over {V_{tb} V_{tq}^*}}(10/9
+G(m_c,\mu,q^2))]$ and
$P_e = (\alpha_{em}/9\pi)(3\bar c_1+\bar c_2) [{{V_{cb} V_{cq}^*}\over {V_{tb}
V_{tq}^*}}(10/9 + G(m_c,\mu,q^2))+{{{V_{ub} V_{uq}^*}\over {V_{tb}
V_{tq}^*}}}(10/9 + G(m_c,\mu,q^2))]$. $V_{i,j}$ are the elements of the CKM
matrix.
$m_c$ is the charm quark mass and $m_u$ is the up quark mass. The function
$G(m,\mu,q^2)$ is give by
\begin{eqnarray} G(m,\mu,q^2) = 4\int^1_0 x(1-x) \mbox{d}x
\mbox{ln}{m^2-x(1-x)q^2\over
\mu^2}\;.
\end{eqnarray} In the numerical calculation, we will use $q^2 = m_b^2/2$ which
represents the average value and the full expressions for $P_{s,e}$. In Table 1
we show the values of the effective Wilson coefficients at the scale $m_b$ and
$m_b/2$ for the process $b\rightarrow sq\bar q$. Values for $b\rightarrow dq\bar
q$ can be similarly obtained. These co-efficients are scheme independent and
gauge invariant.

\section{Matrix elements in factorization approximation}  The generalized
factorizable approximation has been quite successfully used in two body D decays
as well as $B\rightarrow D$ decays. The method includes color octet non
factorizable contribution by treating $\xi\equiv 1/N_c$ as an adjustable
parameter \cite{desh}. Justification for this process has recently discussed from
QCD considerations\cite{neubert,stech2}. In general $\xi$ is process dependent,
but using SU(3) flavor symmetry, it should be the same for
$B\rightarrow \pi\pi,\,K\pi,\, K\eta^{\prime}(\eta)$ system. Establishing the
range of value of $\xi$ for the best fit will be one of our goals.

Technique of parametrizing a two body decay amplitude in factorization
approximation is well known. Here we shall do it for $B\rightarrow
K\eta^{\prime}(\eta)$ process to establish our notation and discuss some special
issues relating to this process.  

We define the decay constants and  the form factors as 
\begin{eqnarray} <0|A_{\mu}|M(p)> &=& i f_{M} p_{\mu},  \\ 
<M(p^{\prime})|V_{\mu}|B(p)> 
 &=& \left[(p^{\prime}+p)_{\mu} - {m^2_B-m^2_K \over q^2} q_{\mu} \right]
F^{B\rightarrow M}_1(q^2) 
  + {m^2_B-m^2_K \over q^2} q_{\mu} F^{B\rightarrow M}_0(q^2) , 
\end{eqnarray}  where $M$, $V_{\mu}$ and $A_{\mu}$ denote a pseudoscalar meson,
a vector current and an  axial-vector current, respectively, and
$q=p-p^{\prime}$.   Note that $F_1(0)=F_0(0)$ and we can set 
$F^{B\rightarrow M}_{0,1}(q^2=m^2_M)
\approx F^{B\rightarrow M}_{0,1}(0)$ since these form factors are pole dominated
by mesons at scale $m^2_B$.  

The physical states $\eta$ and $\eta^{\prime}$ are mixtures of SU(3) singlet
state $\eta_1$  and octet state $\eta_8$ : 
\begin{eqnarray}
\eta = \eta_8 \cos\theta -\eta_1 \sin\theta,  \ \ \ \ 
\eta^{\prime} = \eta_8 \sin\theta +\eta_1 \cos\theta,
\end{eqnarray}  with 
\begin{eqnarray}
\eta_8 = {1 \over \sqrt{6}} ( u \bar u+ d \bar d -2 s \bar s), \ \ \ \
\eta_1 = {1 \over \sqrt{3}} ( u \bar u+ d \bar d + s \bar s) , 
\end{eqnarray}    The decay constants $f^u_{\eta}$ and $f^s_{\eta}$, which are
similarly defined as Eqs.(\ref{eta1})  and (\ref{eta2}), have the relations
similar to Eq.(\ref{feta}): 
\begin{eqnarray} f^u_{\eta} ={f_8 \over \sqrt{6}} \cos\theta -{f_1 \over
\sqrt{3}} \sin\theta, \ \ \ \ f^s_{\eta} =-2 {f_8 \over \sqrt{6}} \cos\theta
-{f_1 \over \sqrt{3}} \sin\theta.  
\end{eqnarray}  In SU(3) limit, $f_K =f_{\pi}=f_8$. However from light quark
meson decays their values can be obtained. In particular the values of $f_8$ and
$f_1$ can be obtained from
 $\eta\rightarrow\gamma\gamma$ and $\eta^{\prime}\rightarrow\gamma\gamma$
provided the mixing angle $\theta$ is known. We shall see later that larger
magnitude of $\theta$ enhances the $\eta^{\prime}$ decays. We shall thus use the
value $\theta=$- 25$^0$ which leads to $f_8\sim 1.75 f_{\pi}$ and $f_8\sim
f_{\pi}$ \cite{F} and we use $f_{\pi}=132$ MeV and $f_K$=158 MeV. A technical
point is to note that when we  evaluate $<0|\bar si\gamma_5 s|\eta>$
or
$<0|\bar si\gamma_5 s|\eta^{\prime}>$, because of anomalies in the corresponding
current $\bar s\gamma_{\mu}\gamma_{5}s$, we use anomaly free currents and
neglect terms corresponding to light quark masses as discussed in ref\cite{F}.
We then have 
\begin{eqnarray} <0|\bar si\gamma_5 s|\eta^{\prime}>&=&-{\sqrt 3\over\sqrt
2}{{f_8\sin\theta m^2_{\eta^{\prime}}}\over{2 m_s}}\\\nonumber <0|\bar si\gamma_5
s|\eta>&=&-{\sqrt 3\over\sqrt 2}{{f_8\cos\theta m^2_{\eta}}\over{2 m_s}}
\end{eqnarray}
 The decay constants $f^u_{\eta^{\prime}}$ and
$f^s_{\eta^{\prime}}$ are defined as
\begin{eqnarray}
\label{eta1} <0|\bar u \gamma_{\mu} \gamma_5 u|\eta^{\prime}> = i
f^u_{\eta^{\prime}} p_{\mu}, \\
\label{eta2} <0|\bar s \gamma_{\mu} \gamma_5 s|\eta^{\prime}> = i
f^s_{\eta^{\prime}} p_{\mu}.
\end{eqnarray}  Due to $\eta - \eta^{\prime}$ mixing, $f^u_{\eta^{\prime}}$ and
$f^s_{\eta^{\prime}}$ are  related to $f_8$ and $f_1$ by 
\begin{eqnarray}
\label{feta} f^u_{\eta^{\prime}} ={f_8 \over \sqrt{6}} \sin\theta +{f_1 \over
\sqrt{3}} \cos\theta, \ \ \ \ f^s_{\eta^{\prime}} =-2 {f_8 \over \sqrt{6}}
\sin\theta +{f_1 \over \sqrt{3}} \cos\theta,  
\end{eqnarray}  where $f_8$ and $f_1$ are defined as 
\begin{eqnarray} <0|\bar u \gamma_{\mu} \gamma_5 u|\eta_8> = i {f_8 \over
\sqrt{6}} p_{\mu}, \ \ \ \  <0|\bar u \gamma_{\mu} \gamma_5 u|\eta_1> = i {f_1
\over \sqrt{3}} p_{\mu}. 
\end{eqnarray}  

We shall assume that form factors are related by nonet symmetry. For a current
$V_\mu={\bar u}\gamma_{\mu}b$ this implies:
\begin{eqnarray} F^{B\rightarrow K}_{0}& =&F^{B\rightarrow\pi^{-}}_{0}
\\\nonumber &=&\sqrt{2}F^{B\rightarrow \pi^{0}}_{0}\\\nonumber
&=&\sqrt{6}F^{B\rightarrow \eta_{8}}_{0}\\\nonumber &=&\sqrt{3}F^{B\rightarrow
\eta_{0}}_{0}
\end{eqnarray}  We expect SU(3) breaking effect could be $O(15)\%$. In
particular 
$F^{B\rightarrow \eta_{0}}_{0}$ could be smaller if $\eta_0$ has significant
glue content. Form factors $F^{B\rightarrow \eta}_{0}$ and $F^{B\rightarrow
\eta^{\prime}}_{0}$ are then
 
\begin{eqnarray} F^{B\rightarrow \eta}_{0} &=&F^{B\rightarrow \pi^-}_{0}
(\cos\theta / \sqrt{6} - \sin\theta /
\sqrt{3})\\\nonumber
 F^{B\rightarrow \eta^{\prime}}_{0} &=&F^{B\rightarrow \pi^-}_{0} (\sin\theta /
\sqrt{6} + \cos\theta /
\sqrt{3}).
\end{eqnarray}

There seems to be considerable variation in the range of
$F_0^{B\rightarrow\pi^-}$ estimated in the literature . Bauer et al. \cite{stech}
estimates it at 0.33 while Deandrea et al. \cite{gato} obtains 0.5. Since the
rate is proportional to the
$|F_0|^2$, this can be a source of considerable error. We find that data on
$B\rightarrow\pi^+\pi^-$ mode places rather stringent constraint on the
magnitude of the form factors with values  near  Bauer et al. being preferred.  
  
The decay amplitude  for 
$B^- \rightarrow \eta^{\prime} K^-$ is now found to be: 
\begin{eqnarray}
\label{AA} A(B^- \rightarrow \eta^{\prime} K^-) &=& {G_{F} \over \sqrt{2}}
    \{ V_{ub} V^{*}_{us} [(c_1 +\xi c_2) C^u + (\xi c_1 +c_2) T]  \nonumber \\ 
  & & \mbox{} - V_{tb} V^{*}_{ts} [(c_3 +\xi c_4-c_5-\xi c_6) (2 C^u+C^s) 
              +(\xi c_3+c_4) (C^s+T)                 \nonumber \\ 
  & & \mbox{} +2 (\xi c_5 +c_6)(X {\bar C}^s+Y T) 
              -{1 \over 2} (c_7 +\xi c_8 -c_9 -\xi c_{10}) (C^u-C^s)   
\nonumber \\ 
  & & \mbox{} -(\xi c_7 +c_8) X {\bar C}^s+2 (\xi c_7+c_8) Y T              
              -{1 \over 2} (\xi c_9 +c_{10}) C^s +(\xi c_9+c_{10}) T]
\}\nonumber\\+A_{11}, 
\end{eqnarray}  where 
\begin{eqnarray}
\label{CCTXY}
 C^u &=& i f^u_{\eta^{\prime}} F^{B\rightarrow K}_0 (m^2_{B} - m^2_{K}),
\nonumber \\  
 C^s &=& i f^s_{\eta^{\prime}} F^{B\rightarrow K}_0 (m^2_{B} - m^2_{K}),
\nonumber \\ {\bar C}^s&=&-i{\sqrt 3\over\sqrt 2}f_8sin\theta F^{B\rightarrow
K}_0 (m^2_{B} - m^2_{K}),
\nonumber \\  
 T &=& i f_K F^{B\rightarrow  \eta^{\prime}}_0(m^2_{B} - m^2_{\eta^{\prime}}), 
\nonumber \\ 
 X &=& {m^2_{\eta^{\prime}} \over 2 m_s (m_b-m_s)} ,      \nonumber \\ 
 Y &=& {m^2_K \over (m_s+m_u) (m_b-m_u)} .
\end{eqnarray}  Here we have neglected small contribution of the annihilation
term which is proportional to $f_B$. $A_{11}$ represents contribution from
chromomagnetic operator $O_{11}$, and is evaluated as in ref.\cite{a,hdt}. 

The amplitude for $B\rightarrow \eta K$ can be deduced by appropriate replacement
of $\eta^{\prime}$ by $\eta$. In amplitudes where penguin contributions
dominate, we observe that X and Y contributions are very sensitive to the value
of light quark contributions $m_s$. Depending on the scale $\mu$, we have to
employ the corresponding value of $m_s$. We show a plot of $m_s$  as a function
of $\mu$ in Fig.1. We use $m_s$=165 GeV at $\mu=1$ GeV. This leads to
$m_s(m_b/2)=121$ MeV and $m_s(m_b)=118$ MeV. If a smaller value of $m_s(1
GeV)$ is used, process involving K mesons are enhanced. Although this will
enhance $B\rightarrow K\eta^{\prime}$, we will then have too large a value for
$B\rightarrow K^+\pi^-$. We find choice of 165 MeV is optimal. We shall show
later that the
$\mu$ dependence of the rate is quite weak because of the compensating effect of
$\mu$ dependence of $m_s$ and $\mu$ dependence of WC's. Ali and Greub have
advocated that $\eta^\prime$ and $\eta$ might contain a considerable amount of
$c\bar c$ contribution, and this enhances $B\rightarrow
\eta^{\prime} K$. They have argued that if
\begin{eqnarray} <0|{\bar c}\gamma_\mu\gamma_5
c|\eta^\prime(p)>&=&if^c_{\eta^{\prime}}p_\mu\\\nonumber <0|{\bar
c}\gamma_\mu\gamma_5 c|\eta(p)>&=&if^c_{\eta}p_\mu,
\end{eqnarray} then $f^c_{\eta^{\prime}}\simeq 6$ MeV and $f^c_{\eta}\simeq 2.3$
MeV. This should be compared to
$f^u_{\eta^{\prime}}=$50 MeV and $f^u_\eta=$100 MeV. We shall show that it is
possible to fit data without the charm content within 1$\sigma$ of the
experimental error. If further experiments were to narrow the rate for
$B\rightarrow \eta^{\prime} K$ at the upper end of the present range, this would
be a strong argument for the charm content. With the inclusion of charm, the
amplitude in Eq.(\ref{AA}) has to include the term 
\begin{equation} A^{\prime}=-{G_{F} \over \sqrt{2}} V_{cb} V^{*}_{cs}(c_1+\xi
c_2) (f^c_{\eta{\prime}} /f^u_{\eta^{\prime}}) C^u.
\end{equation} For $B\rightarrow\eta K$ we must include a similar term with
$f^c_{\eta^{\prime}}$ replaced by $f^c_{\eta}$.

\section{Decays of B into pseudoscalars}
\subsection{Process $B\rightarrow \pi\pi$}

Here we consider the decays $B^{\pm}\rightarrow\pi^{\pm}\pi^0$, $B^0( \bar
B^0)\rightarrow \pi^{+}\pi^{-}$ and $B^0( \bar B^0)\rightarrow \pi^{0}\pi^{0}$.
Recent measurement at CLEO \cite{cleo1} yield the following
 bound at 90 $\%$ C.L.:
\begin{eqnarray} BR(B^\pm \rightarrow \pi^0 \pi^\pm)< 2\times 10^{-5},
\nonumber\\
 BR(B^0 \rightarrow \pi^+ \pi^-)< 1.5\times 10^{-5}. \nonumber
\end{eqnarray} The decay rates scale as
$|F^{B\rightarrow \pi}_0(0)|^2$ and since the tree diagram dominates the
processes
$B^{\pm}\rightarrow\pi^{\pm}\pi^0$ and $B^0( \bar B^0)\rightarrow
\pi^{+}\pi^{-}$, these two decays also scale as
$|V_{ub}/V_{cb}|^2$. Dependence on the QCD scale $\mu$ for these two decays is
rather mild because the tree amplitude depends on the Wilson coefficients $c_1$
and $c_2$, and these are weakly dependent on $\mu$. Further, the light quark
masses in the matrix elements also scale with $\mu$, partially offsetting the
$\mu$ dependence  from $c_{1,2}$. The partial width for $B^+\rightarrow
\pi^+\pi^0$ for example is obtained from :
\begin{equation}
\Gamma(B^+\rightarrow \pi^+\pi^0)={1\over {8\pi}}{|p|\over
m^2_B}|A(B^+\rightarrow \pi^+\pi^0)|^2
\end{equation}
 where $|p|$ is the pion momentum and the branching ratio is calculated by
multiplying by the total rate
$\tau_B=1.49$ ps. In figures 2-4 we plot branching ratios averaged over particle
and antiparticle  for the modes
$\pi^{\pm}\pi^{0}$,
$\pi^{+}\pi^{-}$  and $\pi^{0}\pi^{0}$, as a function of $\xi\equiv 1/N_c$ for
two different values of the scale $\mu$, $\mu=m_b$ and $\mu=m_b/2$. We have
assumed $|V_{ub}/V_{cb}|=0.07$, $\gamma =35^0$ and the form factor
$F^{B\rightarrow \pi^-}_0=0.36$. We see the weak dependence on the scale $\mu$,
but strong dependence on $\xi$. We shall see later that to enhance  $B\rightarrow
\eta^{\prime}K$ values of $\xi\sim 0$ are preferred. In the range where
$\xi$ is small, the present bounds on the $\pi^+\pi^-$ branching ratio of
1.5$\times 10^{-5}$ already tells us that the product $|{{V_{ub}F}\over
V_{cb}}|\le 0.024$. To enhance
$B\rightarrow \eta^{\prime}K$ a large form factor is preferred. Since $
|V_{ub}/V_{cb}|=0.08\pm 0.02$, we see that we are forced into region of small
$|V_{ub}|$ if we wish to explain $B\rightarrow \eta^{\prime}K$ without invoking
new physics. Further, the form factor can not be taken larger than 0.4 without
violating the present bounds on $|V_{ub}/V_{cb}|$. The value of $\gamma$ used
does not alter the above conclusions, it however will be important when we
consider CP violating effects. The ratio of $B^{\pm}\rightarrow\pi^{\pm}\pi^0$
and $B^0( \bar B^0)\rightarrow \pi^{+}\pi^{-}$ is not sensitive to the form
factor or $V_{ub}$, but is very sensitive to values of
$\xi$. In fig.5 we plot this ratio for
$\mu=m_b$. Future measurements of this ratio will constrain the value of
$\xi$. We shall see later that larger form factor, although favorable in
increasing 
$B\rightarrow \eta^{\prime}K$, also enhances $B\rightarrow \pi K$, resulting in
conflict with experiment. We find that the form factor $|F^{B\rightarrow
\pi^-}_0|=0.36$ and
$|V_{ub}/V_{cb}|\sim 0.07$ are the best compromise. In summary, bounds on
$B\rightarrow \pi^+\pi^-$ already provide a strong constraints on the size of
the form factors and the value of $|V_{ub}/V_{cb}|$.

\subsection{Processes $B\rightarrow \pi K$, $B^{\pm}\rightarrow
\eta^{\prime}K^{\pm}$ and $B^{\pm}\rightarrow \eta K^{\pm}$} 

We now examine the two body processes involving kaons. Recent measurement at
CLEO \cite{{cleo0},{cleo1}} yield the following  bound: 
\begin{eqnarray} BR(B^\pm \rightarrow \pi^\pm K)=  (2.3^{+1.1+0.2}_{-1.0-0.2}\pm
0.2)\times 10^{-5} \;\; \;, 
 \nonumber\\  BR(B^0 \rightarrow \pi^\pm K^{\mp})=  (1.5^{+0.5+0.1}_{-0.4-0.1}\pm
0.1)\times 10^{-5} \;\; \;, \nonumber
\\ \nonumber BR(B^{\pm} \rightarrow \eta^{\prime} K^{\pm})= 
(7.8^{+2.7}_{-2.2}\pm 1.0)\times 10^{-5} \;\; \;. \nonumber
\end{eqnarray} We again choose the value of the form factor
$F^{B\rightarrow K}_{0}=0.36$, and weak phase
$\gamma\sim 35^0$. In fig.6 and 7, we have plotted the branching ratio for
$B^+\rightarrow \pi^+ K^0$ and $B^0\rightarrow \pi^-K^+$, averaged over
particle-antiparticle decays as a function of $\xi$ for $\mu=m_b$ and
$\mu=m_b/2$. There is only a slight $\mu$ dependence with $B^+\rightarrow
\pi^+K^0$ being slightly larger for $\mu=m_b$. Both rates are enhanced at small
$\xi$. In particular the observed branching ratio of $B^0\rightarrow \pi^-K^+$
already  constrain $\xi>0.1$. If a smaller value of form factor is used, the
rate for
$B\rightarrow \eta^{\prime}K$ will go down correspondingly. We have also plotted
the average value of $B^0\rightarrow \pi^0 K^0$ as a function of $\xi$ for two
different $\mu$ in fig.8.

Turning to $B\rightarrow \eta^{\prime} K$, we first examine the $\xi$ dependence
for two different values of $\mu$. We again see in fig.9 an enhancement for small
$\xi$ and slightly larger values for $\mu=m_b$. This figure is based on
$\eta-\eta^{\prime}$ mixing of $\theta=-25^0$. Clearly, values of $\xi=0.2$ is
consistent with data at 1$\sigma$. It is not possible to enhance the rate by
increasing the form factor, because $B^0\rightarrow \pi^-K^+$ will then become
too large.

 We have examined the branching ratio dependence of
$B\rightarrow \eta^{\prime} K$ on the mixing angle $\theta$. In fig.10 we plot
the branching ratio for
$B\rightarrow \eta^{\prime} K$ as a function of $\theta$, and find that the
branching ratio increases as $\theta$ becomes more negative. 

From experiment we also have the following bound at 1$\sigma$:
\begin{equation} R={{BR(B^+\rightarrow \eta^{\prime}K^+)}\over
{BR(B^0\rightarrow \pi^- K^+)}}\geq 2.7
\end{equation} In fig.11, we plot this ratio as a function of the weak phase
$\gamma$ for
$\xi=0.1$. Since this ratio does not depend on the size of the form factors, or
the value of the $|V_{cb}|$, we see that there is strong preference for the
values of small
$\gamma$. We therefore have chosen a small value of $\gamma\sim 35^0$ for our
plots.
   
 If further experiments reduce the error on the $B\rightarrow \eta^{\prime}K$
branching ratio, and it turns out to be a larger number, one may have to consider
Ali-Greub suggestion of including the charm content. With values of
$f^c_{\eta^{\prime}}\simeq 5.8$  MeV, and sign so chosen to give a constructive
interference, we plot the branching ratio of $B\rightarrow K\eta^{\prime}$ as a
function of $\xi$ in fig.12. As we see, there is about a 15$\%$ enhancement in
the rate at
$\xi=0$.

We consider $B\rightarrow\eta K$ as a function of $\xi$ with or without
inclusion of charm in fig.13 For $\xi=0.1$, the branching ratio is of the order
of $5\times 10^{-6}$ and the process will not be hard to observe. Inclusion of
electroweak penguin contribution actually suppress this decay significantly. We
show in fig. 14, the branching ratio without electroweak penguin and with
electroweak penguin.

\section{CP asymmetry in the decay modes} So far we have found that the
branching ratio of $B^+\rightarrow
\eta^{\prime}K^+$ is large if we go to a parameter space where $\xi$ is small,
form factor is large, weak phase $\gamma$ is small and the $\eta-\eta^{\prime}$
mixing angle
$\theta$ is $\simeq -25^0$. We now calculate the rate asymmetry for $B\rightarrow
\eta^{\prime}K$, $B\rightarrow \eta K$ and $B^0\rightarrow \pi^-K^+$ mode in this
parameter space. Interestingly enough we find that the rate asymmetry to be
10$\%$ for the  $B\rightarrow \eta K$ when $\gamma$ is around 110$^0$ and
$\xi=0.1$. In fig.15 we plot the rate asymmetry for the $B\rightarrow \eta K$
against different values of $\gamma$ for a fixed $\xi=0.1$. If we include the
charm content the rate asymmetry is slightly higher or lower depending on the
sign of $f^c_{\eta}$. In the fig.16, we show the rate asymmetry for
$B\rightarrow \eta^{\prime}K$ as a function of $\gamma$. The asymmetry in the
$B\rightarrow \eta^{\prime}K$ mode is largest about 2$\%$ when $\gamma$ is large
$\sim$ 85$^0$. In the fig.17 we show the rate asymmetry for the $B^0\rightarrow
\pi^-K^+$ mode as a function of $\gamma$. For $\gamma$ of about 35$^0$ and
$\xi=0.2$ the rate asymmetry in this mode is about 5$\%$. The asymmetry
maximizes for $\gamma$ around 70$^0$ for $\xi=0.2$ 

\section{ Conclusion }

We have shown that it is possible to understand the decays of B meson to light
pseudoscalar mesons, i.e. $\pi\pi$, $\pi K$ and $\eta K$, $\eta^{\prime} K$
within factorization approximation. No new physics is needed.  We can have large
branching ratio for $B\rightarrow \eta^{\prime}K$ as seen by CLEO, in the
parameter space which is not excluded by the other experimental observations.
This region is found for $0.1<\xi<0.2$. The parameters which we have varied to
fit all the data are the  form factors, the QCD scale ($\mu$), 
$\xi(\equiv 1/N_c)$, the absolute values of the CKM elements, the weak phases and
the
$\eta-\eta^{\prime}$ mixing angle $\theta$. We found that the  large form factor
helps to increase the branching ratio of $B\rightarrow
\eta^{\prime}K$. But $B\rightarrow\pi\pi$ and $B\rightarrow \pi K$ decays
restrict the size of these form factors.  We also have found that smaller values
of
$\xi$ enhances the branching ratio of $B\rightarrow\eta^{\prime} K$. In order to
find the dependence on $\gamma$, we have studied the ratio of the branching ratio
of
$B\rightarrow\eta^{\prime} K$ and the branching ratio of
$B\rightarrow\pi K$. The ratio does not depend on the form factors and we have
found that the small value of the weak phase $\gamma\simeq 35^0$ is preferred. We
have found that the smaller
$|V_{ub}/V_{cb}|$ is preferred. Our investigation is closest in spirit to Ali
and Greub \cite{c}. They choose the QCD scale $\mu=m_b/2$. We have found that
$\mu$ dependence introduces only a small effect on decay rates. We have included
electroweak penguin contributions, these are important for $B\rightarrow\eta K$.
We have examined dependence on mixing $\theta$ and the weak phase $\gamma$. We
agree on preference for small $\gamma$. We also agree that small values of $\xi$
are preferred. We do not find the need for charm in $\eta^{\prime}$ compelling.
We have also looked at CP asymmetries in the allowed parameter space and have
found two modes where it may be possible to measure this asymmetry in future. 
The CP asymmetries for these two modes are   (i) $B^{\pm}\rightarrow\eta K^{\pm}
$ about 10
$\%$ and (ii) $B^{0}\rightarrow\pi^{\pm} K^{\mp} $ about 5$\%$.  

This work was supported in part by the US Department of Energy Grants No. 
DE-FG06-854ER-40224 and DE-FG03-95ER40894.

\newpage
\noindent {\bf Figure Captions:}

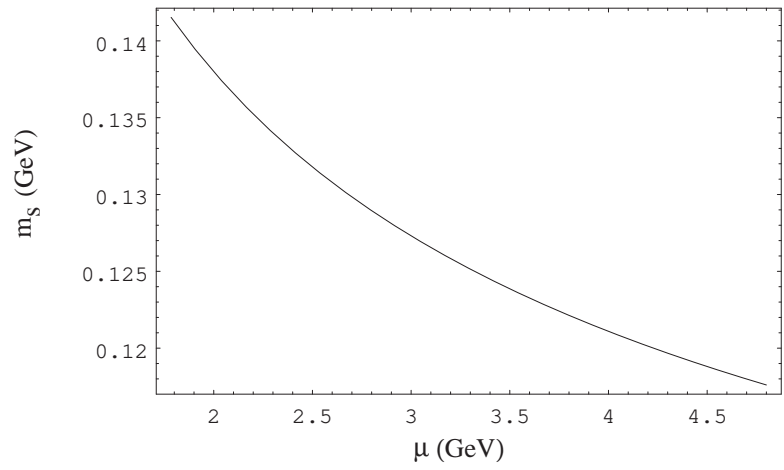
\noindent {\bf Fig. 1}: Strange quark mass $m_s$ is plotted as a function of
$\mu$. $\mu$ has been varied between the  $m_\tau$ mass and  the $m_b$ mass.

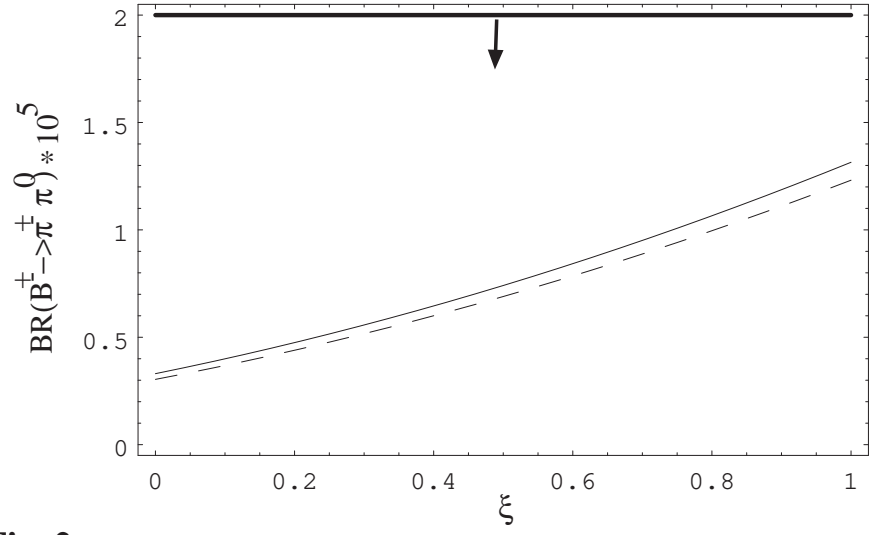
\noindent {\bf Fig. 2}: Branching ratio for the average of
$B^{\pm}\rightarrow\pi^{\pm}\pi^0$ as a function of $\xi$. The solid curve is for
$\mu=m_b/2$ and the dashed curve is for $\mu=m_b$. the direction of the thick
arrows indicate the regions being allowed by the available experimental data.

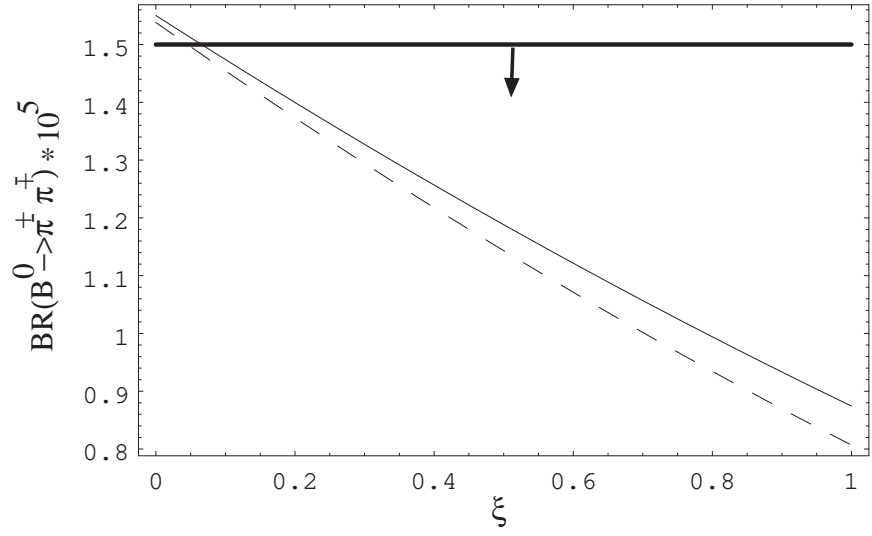
\noindent {\bf Fig. 3}: Branching ratio for the  average of
$B^{0}({\bar B^{0}})\rightarrow\pi^{+}\pi^-$ as a function of $\xi$. The solid
curve is for
$\mu=m_b/2$ and the dashed curve is for
$\mu=m_b$.

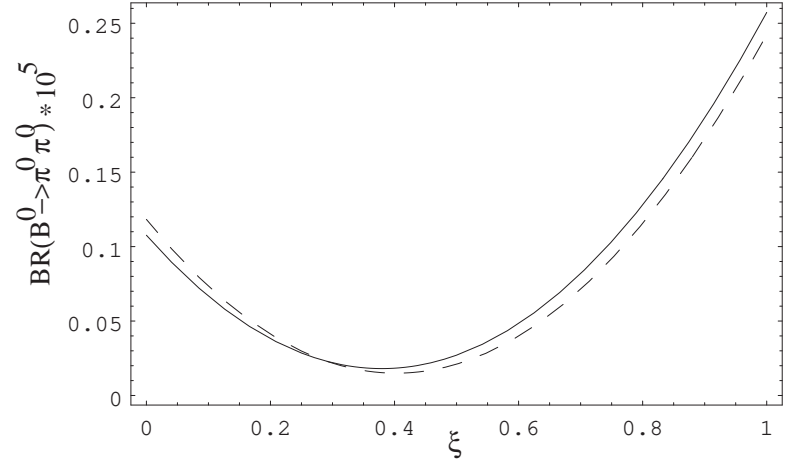
\noindent {\bf Fig. 4}: Branching ratio for the  average of $B^{0}({\bar
B^{0}})\rightarrow\pi^{0}\pi^0$ as a function of $\xi$. The solid curve is for
$\mu=m_b/2$ and the dashed curve is for
$\mu=m_b$.

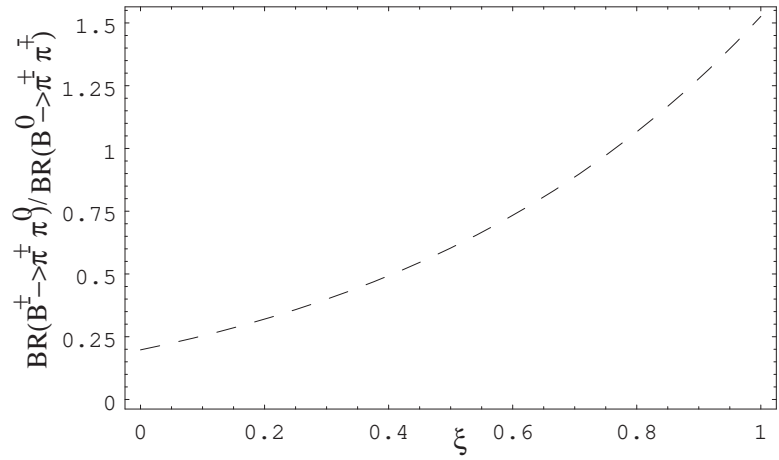
\noindent {\bf Fig. 5}: Ratio of the branching ratio of 
$B^{\pm}\rightarrow\pi^{\pm}\pi^0$ and
$B^{0}({\bar B^{0}})\rightarrow\pi^{+}\pi^{-}$ as a function of $\xi$. The curve
is drawn for
$\mu=m_b$.

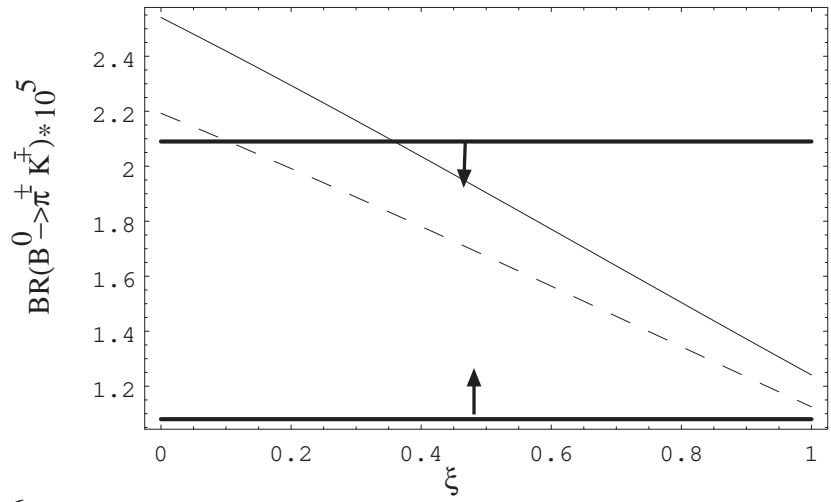
\noindent {\bf Fig. 6}: Branching ratio for the average of
$B^{0}({\bar B^{0}})\rightarrow\pi^{\pm}K^{\mp}$ as a function of $\xi$. The
solid curve is for
$\mu=m_b/2$ and the dashed curve is for $\mu=m_b$.

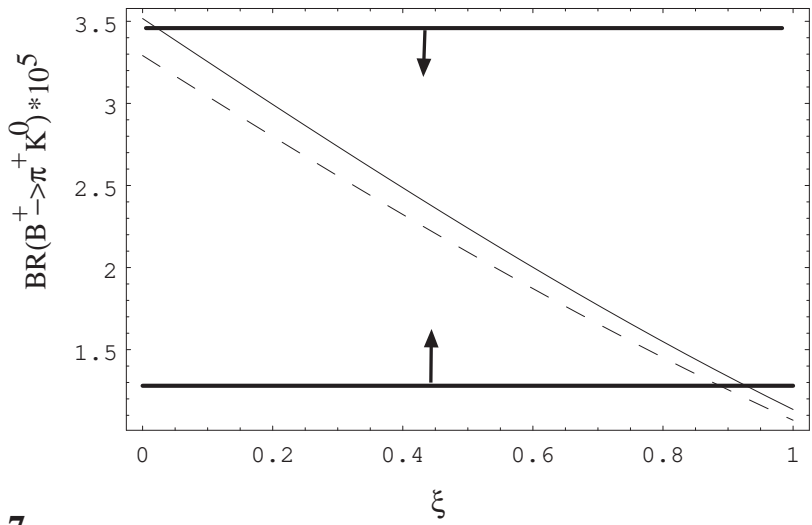
\noindent {\bf Fig. 7}: Branching ratio for the average of
$B^{\pm}\rightarrow\pi^{\pm}K^{0}$ as a function of $\xi$. The solid curve is for
$\mu=m_b/2$ and the dashed curve is for $\mu=m_b$. 

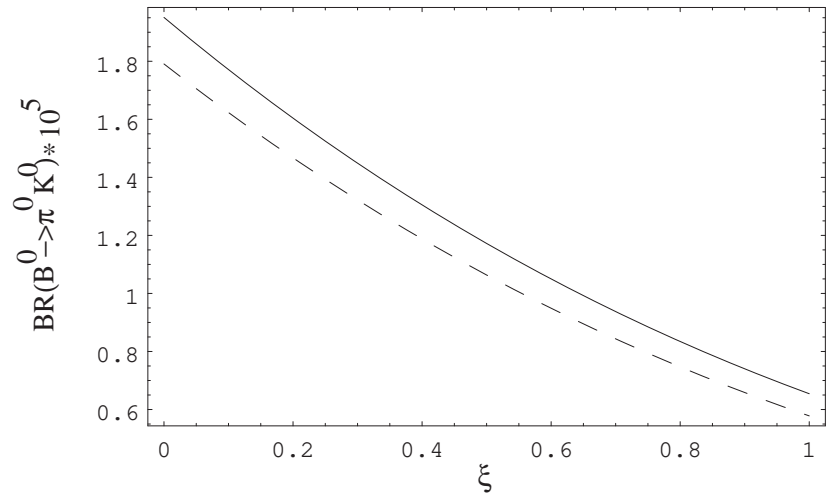
\noindent {\bf Fig. 8}: Branching ratio for the average of
$B^{0}({\bar B^{0}})\rightarrow\pi^{0}K^{0}$ as a function of $\xi$. The solid
curve is for
$\mu=m_b/2$ and the dashed curve is for $\mu=m_b$.  

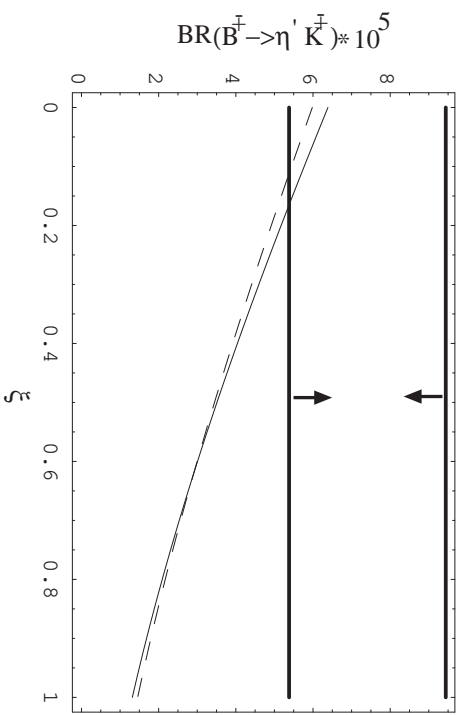
\noindent {\bf Fig. 9}: Branching ratio for the average of
$B^{\pm}\rightarrow\eta^{\prime}K^{\pm}$ as a function of $\xi$. The solid curve
is for $\mu=m_b/2$ and the dashed curve is for $\mu=m_b$. 

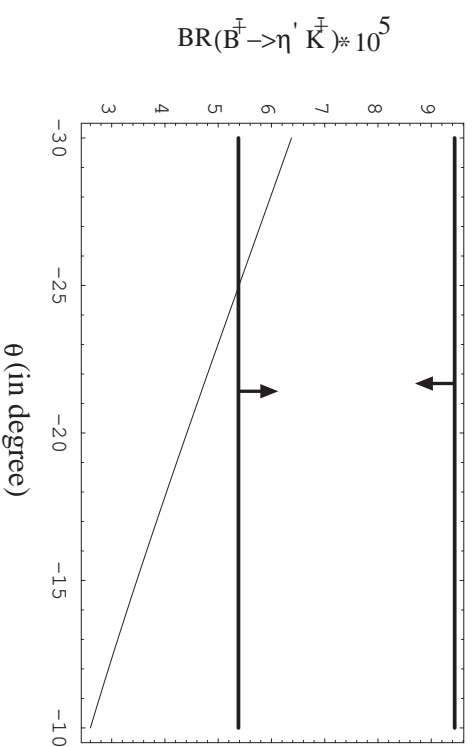
\noindent {\bf Fig. 10}: Branching ratio for the average of
$B^{\pm}\rightarrow\eta^{\prime}K^{\pm}$ as a function of $\theta$. The curve is
drawn at $\mu=m_b$. 

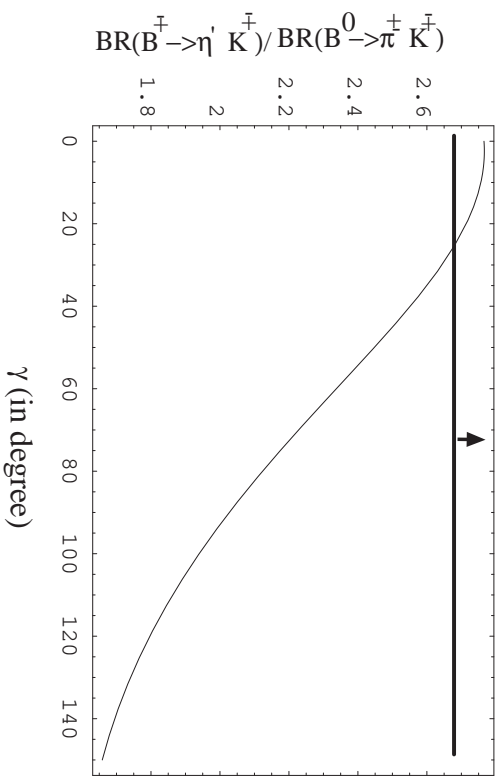
\noindent {\bf Fig. 11}: Ratio of the branching ratio for
$B^{\pm}\rightarrow\eta^{\prime}K^{\pm}$ and
$B^{0}({\bar B^{0}})\rightarrow\pi^{\pm}K^{\mp}$ as a function of $\gamma$. The
curve is drawn at $\mu=m_b$.

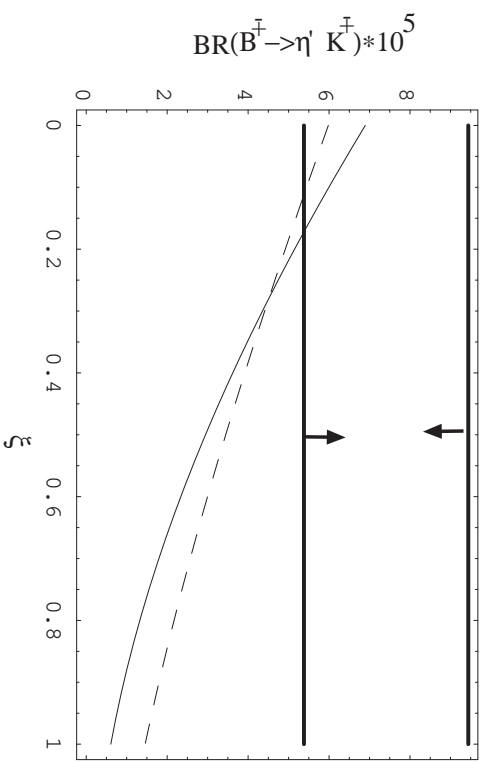
\noindent {\bf Fig. 12}: Branching ratio for the average of
$B^{\pm}\rightarrow\eta^{\prime}K^{\pm}$ as a function of $\xi$ (dashed line).
Same branching ratio but with the assumption that  $\eta^{\prime}$ has charm
content(solid line) with $f^c_{\eta^{\prime}}=5.8$ MeV. Both the lines are drawn
for
$\mu=m_b$.

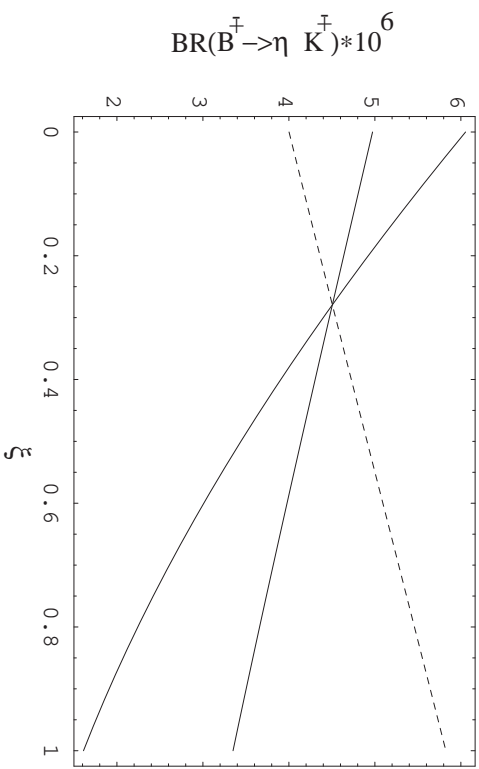
\noindent {\bf Fig. 13}: Branching ratio for the average of
$B^{\pm}\rightarrow\eta K^{\pm}$ as a function of $\xi$ (dashed line). Same
branching ratio but with the assumption that  $\eta$ has charm content (solid
line and small dashed  line). The solid line and the small dashed line have
different combination of signs for  $f^c_{\eta}$. We have used $f^c_{\eta}=2.3$
MeV.  All the contours are drawn for $\mu=m_b$.

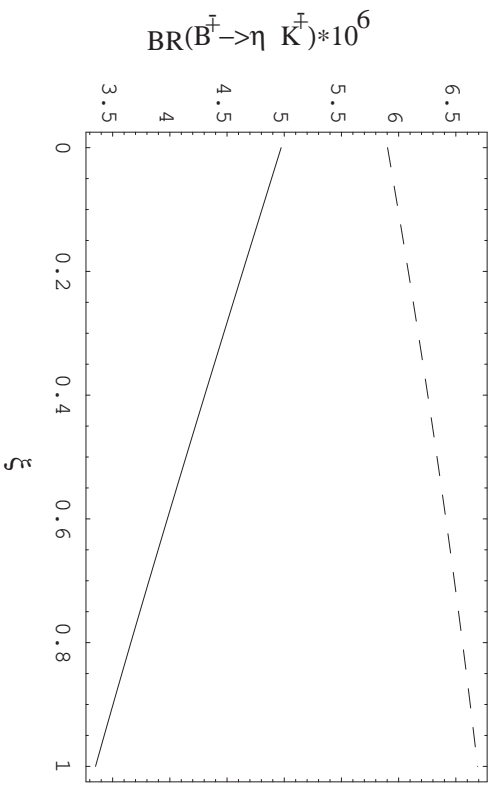
\noindent {\bf Fig. 14}: Branching ratio for the average of
$B^{\pm}\rightarrow\eta K^{\pm}$ as a function of $\xi$ (solid line). Same
branching ratio but without the electroweak contribution (dashed line). Both the
lines are drawn for $\mu=m_b$.
 
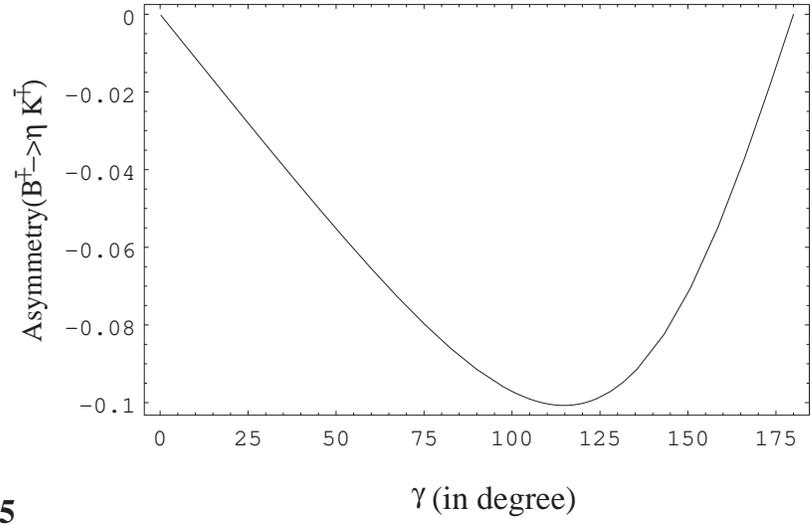
\noindent {\bf Fig. 15}: CP asymmetry for the mode
$B^{\pm}\rightarrow\eta^{\prime}K^{\pm}$ as a function of $\gamma$.

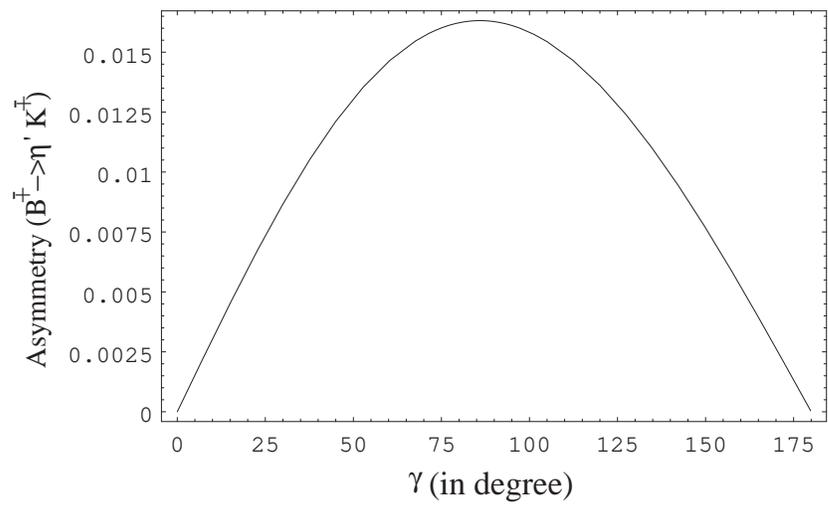
\noindent {\bf Fig. 16}: CP asymmetry for the mode
$B^{\pm}\rightarrow\eta K^{\pm}$ as a function of $\gamma$.

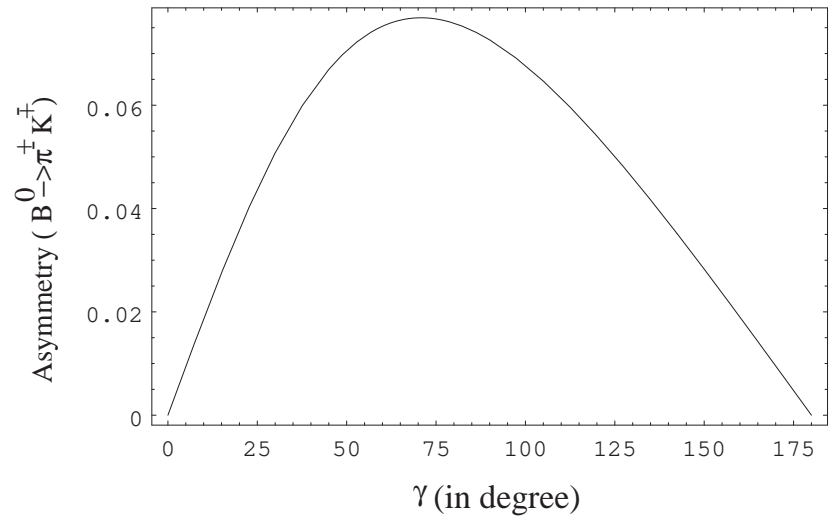
\noindent {\bf Fig. 17}: CP asymmetry for the mode
$B^{0}\rightarrow\pi^{\pm}K^{\mp}$ as a function of $\gamma$.

\noindent {\bf Table Caption:}

\noindent {\bf Table 1}: Effective Wilson coefficients for the $b\rightarrow s$
transition at the scale $m_b$ and $m_b/2$ are shown.

\newpage 
\begin{center}  Table 1 \end{center}
\begin{center}
\begin{tabular}{|c|c|c|}  \hline  WC's&$\mu$=${m_b\over 2}$ &$\mu$=$m_b$\\\hline 
$C^{\rm eff}_1$&-0.282 &-0.3209\\\hline
$C^{\rm eff}_2$&1.135 &1.149\\\hline
$C^{\rm eff}_3$&0.0228718+i0.004689 & 0.02175-i0.0041396 \\\hline
$C^{\rm eff}_4$&-0.051144-i0.004689 & -0.04906 - i0.0124188\\\hline
$C^{\rm eff}_5$&0.0162153+i 0.004689&0.015601+i0.0041396\\\hline
$C^{\rm eff}_6$&-0.0653549-i 0.0140673 & -0.060632 - i0.0124188\\\hline
$C^{\rm eff}_7$& 0.00122773+i0.00005724 & -0.000859 + i0.0000728\\\hline
$C^{\rm eff}_8$&-0.0000953211 &0.00143303\\\hline
$C^{\rm eff}_9$&-0.0120155+i0.0000572433 & -0.011487 + i0.0000727\\\hline
$C^{\rm eff}_{10}$&0.00218628 &0.00317436\\\hline
$C^{\rm eff}_{11}$&-0.334 &-0.295\\\hline
\end{tabular}
\end{center}

\newpage
\begin{figure}[htb]
\vspace{1 cm}

\centerline{ \DESepsf(fig1v2.epsf width 12 cm) }

\centerline{ \DESepsf(fig23.epsf width 12 cm) }

\centerline{ \DESepsf(fig45.epsf width 12 cm) }

\centerline{ \DESepsf(fig67v2.epsf width 12 cm) }

\centerline{ \DESepsf(fig8v2.epsf width 12 cm) }

\centerline{ \DESepsf(fig910v2.epsf width 12 cm) }

\centerline{ \DESepsf(fig1112v2.epsf width 12 cm) }

\centerline{ \DESepsf(fig1314v2.epsf width 12 cm) }

\centerline{ \DESepsf(fig1516v2.epsf width 12 cm) }

\centerline{ \DESepsf(fig17v2.epsf width 12 cm) }

\end{figure}

\end{document}